\documentclass{ws-ijmpcs}

\usepackage{graphicx}
\usepackage{dcolumn}
\usepackage{bm}
\usepackage{amssymb}
\usepackage{amsfonts}
\usepackage{amsmath}
\usepackage{latexsym}
\usepackage{dsfont}
\usepackage{multirow}
\usepackage{color}
\usepackage[mathscr]{eucal}

\newcommand{\beq}{\begin{equation}}
\newcommand{\eeq}{\end{equation}}
\newcommand{\beqy}{\begin{eqnarray}}
\newcommand{\eeqy}{\end{eqnarray}}
\newcommand{\beqyn}{\begin{eqnarray*}}
\newcommand{\eeqyn}{\end{eqnarray*}}

\newcommand{\bs}{\begin{slide}}
\newcommand{\es}{\end{slide}}
\newcommand{\bc}{\begin{center}}
\newcommand{\ec}{\end{center}}
\newcommand{\bmin}{\begin{minipage}}
\newcommand{\emin}{\end{minipage}}

\newcommand{\bea}{\begin{eqnarray}}
\newcommand{\eea}{\end{eqnarray}}
\newcommand{\be}{\begin{equation}}
\newcommand{\ee}{\end{equation}}

\newcommand{\ud}{\mathrm{d}}

\newcommand{\uTr}{\mathrm{Tr}}

\newcommand{\sgn}{\text{sgn}}
\newcommand{\barpsi}{\overline{\psi}}

\newcommand{\uvec}[1]{\boldsymbol{#1}}

\def\trimmarks{}

\begin{document}

\markboth{C. Lorc\'e}
{Quark spin-orbit correlations}

%
\catchline{}{}{}{}{}
%

\title{Quark spin-orbit correlations}

\author{C\'edric Lorc\'e}

\address{IFPA, AGO Department, Universit\'e de Li\`ege, 
  Sart-Tilman, 4000 Li\`ege, Belgium\\
C.Lorce@ulg.ac.be}

\maketitle

\begin{history}
\received{Day Month Year}
\revised{Day Month Year}
\published{Day Month Year}
\end{history}

\begin{abstract}
The proton spin puzzle issue focused the attention on the parton spin and orbital angular momentum contributions to the proton spin. However, a complete characterization of the proton spin structure requires also the knowledge of the parton spin-orbit correlation. We showed that this quantity can be expressed in terms of moments of measurable parton distributions. Using the available phenomenological information about the valence quarks, we concluded that this correlation is negative, meaning that the valence quark spin and kinetic orbital angular momentum are, in average, opposite. The quark spin-orbit correlation can also be expressed more intuitively in terms of relativistic phase-space distributions, which can be seen as the mother distributions of the standard generalized and transverse-momentum dependent parton distributions. We present here for the first time some examples of the general multipole decomposition of these phase-space distributions.
\end{abstract}

\ccode{PACS numbers:11.15.-q,12.38.Aw,13.88.+e,13.60.Hb,14.20.Dh}

\section{Introduction}

Unraveling the spin structure of the nucleon is one of the key questions in hadronic physics. Most of the efforts focused so far on the definition and the determination of the various contributions to the proton spin coming from quarks and gluons\cite{Leader:2013jra,Wakamatsu:2014zza}. We stress that the proton spin structure is actually richer than that.

Because of parity symmetry, the only non-zero correlations involve an even number of angular momenta. For example, the so-called quark orbital angular momentum (OAM) contribution to the proton spin corresponds to twice the correlation between the longitudinal components of the quark OAM $l^q_z$ and the nucleon spin $s^N_z$, namely $L^q_z=2\langle l^q_zs^N_z\rangle$ with $\langle\quad\rangle$ denoting the appropriate average. In these proceedings, we are interested in the quark spin-orbit correlation $C^q_z=2\langle l^q_zs^q_z\rangle$, where $s^q_z$ is the longitudinal component of the quark spin.

\section{Quark orbital angular momentum and spin-orbit correlation}

The kinetic operators associated with quark OAM and spin-orbit correlation are
\begin{align}
\hat L^q_z&=\int\ud^3x\,\tfrac{1}{2}\,\barpsi \gamma^+(\uvec x\times i\overset{\leftrightarrow}{\uvec D}\!\!\!\!\!\phantom{D})_z\psi,\\
\hat C^q_z&=\int\ud^3x\,\tfrac{1}{2}\,\barpsi \gamma^+\gamma_5(\uvec x\times i\overset{\leftrightarrow}{\uvec D}\!\!\!\!\!\phantom{D})_z\psi,\label{SO}
\end{align}
where $\overset{\leftrightarrow}{\uvec D}\!\!\!\!\!\phantom{D}=\overset{\rightarrow}{\uvec\partial}\!\!\!\!\!\phantom{\partial}-\overset{\leftarrow}{\uvec\partial}\!\!\!\!\!\phantom{\partial}-2ig\uvec A$ is the symmetric covariant derivative. The canonical version of these operators is simply obtained \emph{via} the substitution $\uvec D\mapsto\uvec\partial$.

In a famous paper\cite{Ji:1996ek}, Ji has shown that the (kinetic) quark OAM contribution to the proton spin can be expressed in terms of twist-2 generalized parton distributions (GPDs) and form factors (FFs)
\begin{equation}
L^q_z=\tfrac{1}{2}\int\ud x\,x[H_q(x,0,0)+E_q(x,0,0)]-\tfrac{1}{2}\,G^q_A(0).
\end{equation}
Following a somewhat similar approach, we derived\cite{Lorce:2014mxa} the corresponding expression for the (kinetic) quark spin-orbit correlation
\begin{equation}\label{SOtwist2}
C^q_z=\tfrac{1}{2}\int\ud x\,x\tilde H_q(x,0,0)-\tfrac{1}{2}\,[F^q_1(0)-\tfrac{m_q}{2M_N}\,H^q_1(0)].
\end{equation}
At first sight, it may look odd that the spin-orbit correlation $C^q_z$ is related to the helicity GPD $\tilde H_q$ involved in the spin-spin correlation $\langle s^q_zs^N_z\rangle$. However, as discussed by Burkardt\cite{Burkardt:2005hp} in the case of the Ji relation, the extra $x$-factor representing the fraction of longitudinal momentum turns out to provide the ``orbital'' information.

It has also been shown\cite{Penttinen:2000dg,Kiptily:2002nx,Hatta:2012cs} that the (kinetic) quark OAM can alternatively be expressed in terms of twist-3 GPDs
\begin{equation}
L^q_z=-\int\ud x\,xG^q_2(x,0,0).
\end{equation}
We found a similar expression\cite{Lorce:2014mxa} for the (kinetic) quark spin-orbit correlation
\begin{equation}\label{SOtwist3}
C^q_z=-\int\ud x\,x[\tilde G^q_2(x,0,0)+2\tilde G^q_4(x,0,0)].
\end{equation}

Interestingly, the quark OAM and spin-orbit correlation are most intuitively represented in terms of phase-space or Wigner distributions\cite{Lorce:2011kd}. The latter are related by Fourier transform to unintegrated GPDs, also known as generalized transverse-momentum dependent distributions (GTMDs)\cite{Meissner:2009ww,Lorce:2013pza}. In terms of these GTMDs, the quark OAM and spin-orbit correlation read
\begin{align}
L^q_z&=-\int\ud x\,\ud^2k_\perp\,\tfrac{\uvec k^2_\perp}{M^2}\,F^q_{14}(x,0,\uvec k^2_\perp,0,0),\\
C^q_z&=\int\ud x\,\ud^2k_\perp\,\tfrac{\uvec k^2_\perp}{M^2}\,G^q_{11}(x,0,\uvec k^2_\perp,0,0).\label{SOGTMDs}
\end{align}
These expressions have first been obtained for the canonical version of the operators\cite{Lorce:2011kd}. It has been realized later that they remain valid for the kinetic version of the operators, provided that the suitable gauge link is used in the definition of the GTMDs\cite{Ji:2012sj,Lorce:2012ce}. A nice physical interpretation of the difference between canonical and kinetic versions has been proposed by Burkardt\cite{Burkardt:2012sd}. The difference between canonical and kinetic versions of the quark OAM has also been investigated in some models\cite{Lorce:2011kd,Burkardt:2008ua}. While the connection between these GTMDs and experimental observables is not yet clear, recent developments suggest that they could at least be computed on the lattice\cite{Ji:2013dva}.

\section{Phenomenological estimates}\label{sec4}

In the previous section, three different expressions for the quark spin-orbit correlation in terms of parton distributions have been presented. From an experimental perspective, Eq.~\eqref{SOtwist2} is clearly the most useful one. Equating the right-hand side of Eq.~\eqref{SOtwist2} with the right-hand sides of Eqs.~\eqref{SOtwist3} and \eqref{SOGTMDs} leads to two new sum rules among parton distributions.

Following Eq.~\eqref{SOtwist2}, we need to know three quantities in order to determine the quark spin-orbit correlation. We already know the Dirac FF evaluated at $t=0$ in a proton, namely $F^u_1(0)=2$ and $F^d_1(0)=1$. The tensor FF $H^q_1(0)$ can safely be neglected as it comes multiplied by the mass ratio $m_q/4M_N\sim 10^{-3}$ for $u$ and $d$ quarks. So the essential input we need is the second moment of the quark helicity distribution
\beq
\int_{-1}^1\ud x\,x\tilde H_q(x,0,0)=\int_0^1\ud x\,x[\Delta q(x)-\Delta\overline q(x)].
\eeq
Note that contrary to the lowest moment $\int_{-1}^1\ud x\,\tilde H_q(x,0,0)=\int_0^1\ud x\,[\Delta q(x)+\Delta\overline q(x)]$, the second moment cannot be extracted from deep-inelastic scattering (DIS) polarized data without additional assumptions about the polarized sea-quark distributions. However, by combining inclusive and semi-inclusive DIS, one can extract the separate quark and antiquark contributions\cite{Leader:2010rb}. From the LSS fit\cite{Leader:2010rb}, we obtained\cite{Lorce:2014mxa}
\begin{equation}
\int_{-1}^1\ud x\,x\tilde H_u(x,0,0)\approx 0.19,\qquad \int_{-1}^1\ud x\,x\tilde H_d(x,0,0)\approx -0.06,
\end{equation}
at the scale $\mu^2=1$ GeV$^2$, leading to $C^u_z\approx -0.9$ and $C^d_z\approx -0.53$. These values seem consistent with recent Lattice calculations by the LHPC collaboration\cite{Bratt:2010jn}, see table~\ref{Modelresults}.

Since the second moment of the quark helicity distribution is a valence-like quantity with suppressed low-$x$ region, we may expect phenomenological quark model predictions to work better than for the lowest moment. In table~\ref{Modelresults} we provide the first two moments of the $u$ and $d$-quark helicity distributions obtained within the naive quark model (NQM), the light-front constituent quark model (LFCQM) and the light-front chiral quark-soliton model (LF$\chi$QSM)\cite{Lorce:2011dv} at the scale $\mu^2\sim 0.26$ GeV$^2$. From these estimates, we expect a negative quark spin-orbit $C^q_z$ for both $u$ and $d$ quarks ($C^u_z\approx -0.8$ and $C^d_z\approx -0.55$), meaning that the quark spin and kinetic OAM are expected to be, in average, antiparallel. 
\newline

\begin{table}[t]
\tbl{Comparison between the lowest two axial moments for up and down quarks as predicted by the naive quark model (NQM), the light-front constituent quark model (LFCQM) and the light-front chiral quark-soliton model (LF$\chi$QSM) at the scale $\mu^2\sim 0.26$ GeV$^2$, with the corresponding values obtained from the LSS fit to experimental data at $\mu^2=1$ GeV$^2$ and Lattice calculations at $\mu^2=4$ GeV$^2$ and pion mass $m_\pi=293$ MeV.}
{\begin{tabular}{ccccc} \toprule
Model\cite{Lorce:2011dv}&$\int^1_{-1}\ud x\,\tilde H_u(x,0,0)$&$\int^1_{-1}\ud x\,\tilde H_d(x,0,0)$&$\int^1_{-1}\ud x\,x\tilde H_u(x,0,0)$&$\int^1_{-1}\ud x\,x\tilde H_d(x,0,0)$\\
\hline
NQM&$4/3$&$-1/3$&$4/9$&$-1/9$\\
LFCQM&$0.995$&$-0.249$&$0.345$&$-0.086$\\
LF$\chi$QSM&$1.148$&$-0.287$&$0.392$&$-0.098$\\
\hline
LSS\cite{Leader:2010rb}&$0.82$&$-0.45$&$\approx 0.19$&$\approx -0.06$\\
Lattice\cite{Bratt:2010jn}&$0.82(7)$&$-0.41(7)$&$\approx 0.20$&$\approx -0.05$\\
\botrule
\end{tabular}\label{Modelresults}}
\end{table}

\section{Multipole decomposition of Wigner distributions}

The quark GTMD correlator\cite{Meissner:2009ww,Lorce:2013pza}
\begin{equation}\label{GTMDcorr-def}
W^{ab}_{\Lambda'\Lambda}\equiv\int\ud k^-\int\frac{\ud^4z}{(2\pi)^4}\,e^{ik\cdot z}\,\langle P+\tfrac{\Delta}{2},\Lambda'|\overline\psi_b(-\tfrac{z}{2})\mathcal W\psi_a(\tfrac{z}{2})|P-\tfrac{\Delta}{2},\Lambda\rangle,
\end{equation}
with $\mathcal W$ an appropriate Wilson line, is a $2\times 2$ matrix in target polarization space and a $4\times 4$ matrix in Dirac space. In the twist-2 sector, one can interpret
\begin{equation}\label{GTMDcorr}
W_{\vec S\vec S^q}=\tfrac{1}{8}\sum_{\Lambda',\Lambda}(\mathds 1+\vec S\cdot\vec \sigma)_{\Lambda'\Lambda}\,\uTr[W_{\Lambda'\Lambda}\Gamma_{\vec S^q}],
\end{equation}
with $\Gamma_{\vec S^q}=\gamma^++S^q_L\,\gamma^+\gamma_5+S^{qj}_T\,i\sigma^{j+}_T\gamma_5$, 
as the GTMD correlator describing a quark with polarization $\vec S^q$ inside a target with polarization $\vec S$. The corresponding Wigner distributions are obtained by performing the following Fourier transform\cite{Lorce:2011kd}
\begin{equation}
\rho_{\vec S\vec S^q}(x,\uvec k_T,\uvec b_T;\eta)=\int\frac{\ud^2\Delta_T}{(2\pi)^2}\,e^{-i\uvec\Delta_T\cdot\uvec b_T}\,W_{\vec S\vec S^q}(P,k,\Delta;n_+)\big|_{\xi=0},
\end{equation}
where $x=k^+/P^+$ and $\uvec k_T$ are the longitudinal fraction and transverse component of the parton momentum, $\uvec b_T$ is the parton impact parameter, $\xi=-\Delta^+/2P^+$ is the fraction of longitudinal momentum transfer, and $\eta=\sgn(n^0_+)$ with $n_+$ the unit lightlike four-vector satisfying $n^+_+=0$. The hermiticity property of the GTMD correlator~\eqref{GTMDcorr-def} ensures that these Wigner distributions are always real-valued\cite{Lorce:2011ni}, which is in line with their quasi-probabilistic interpretation. 

Playing around with the various polarization configurations, one finds that there are 16 Wigner distributions just like there are 16 possible GTMDs\cite{Meissner:2009ww,Lorce:2013pza}. By construction, the real and imaginary parts of these GTMDs have opposite behavior under naive time-reversal transformation. Similarly, each Wigner distribution can be separated into naive $\mathsf T$-even and $\mathsf T$-odd contributions
\begin{equation}
\rho_{\vec S\vec S^q}=\rho^e_{\vec S\vec S^q}+\rho^o_{\vec S\vec S^q}
\end{equation}
with $\rho^{e,o}_{\vec S\vec S^q}(x,\uvec k_T,\uvec b_T;\eta)=\pm\rho^{e,o}_{\vec S\vec S^q}(x,\uvec k_T,\uvec b_T;-\eta)=\pm\rho^{e,o}_{-\vec S-\vec S^q}(x,-\uvec k_T,\uvec b_T;\eta)$. Using successful phenomenological quark models, we studied the four naive $\mathsf T$-even distributions associated with longitudinal polarization $S^{(q)}_L\equiv\vec S^{(q)}\cdot\vec P/|\vec P|$
\begin{equation}
\rho_{S_LS^q_L}\equiv\rho_{UU}+S_L\,\rho_{LU}+S^q_L\,\rho_{UL}+S_L\,S^q_L\,\rho_{LL},
\end{equation}
and derived the model-independent connection with the OAM\cite{Lorce:2011kd,Lorce:2012ce,Lorce:2011ni,Kanazawa:2014nha}.

The Wigner distributions can be decomposed into two-dimensional multipoles in both $\uvec k_T$ and $\uvec b_T$-spaces. While there is no limit in the multipole order, parity and time-reversal put certain constraints on the allowed multipole. It is therefore more sensible to decompose the Wigner distributions as follows\cite{Lorce:2014}
\begin{equation}
\rho_{\vec S\vec S^q}(x,\uvec k_T,\uvec b_T;\eta)=\sum_m B^m_{\vec S\vec S^q}(\uvec k_T,\uvec b_T;\eta)\,C^m_{\vec S\vec S^q}(x,\uvec k^2_T,(\uvec k_T\cdot\uvec b_T)^2,\uvec b^2_T),
\end{equation}
where $B^m_{\vec S\vec S^q}$ represent the basic multipoles allowed by parity and time-reversal symmetries. These basic multipoles are multiplied by the coefficient functions $C^m_{\vec S\vec S^q}$ which depend on $\mathsf P$ and $\mathsf T$-invariant variables only.

\section{Representation of the transverse phase space}

Wigner distributions are functions of five variables, which are particularly difficult to represent on a two-dimensional sheet of paper. Since we are mainly interested in the transverse phase space, we integrate these distributions over $x$ and set $\eta=+1$. We then represent the transverse Wigner distributions 
\begin{equation}
\rho_{\vec S\vec S^q}(\uvec k_T,\uvec b_T)=\int\ud x\,\rho_{\vec S\vec S^q}(x,\uvec k_T,\uvec b_T;\eta=+1)
\end{equation}
as $\uvec k_T$-distributions at discrete positions in impact-parameter space\cite{Lorce:2014}. 

In the figures presented in this section, we fix $|\uvec b_T|=0.4$ fm and $\phi_b=k\pi/4$ with $k\in\mathbb Z$. The results are obtained in the LFCQM\cite{Lorce:2011dv} for up quarks and normalized to the maximum of the distributions. Light and dark regions represent, respectively, positive and negative domains of the distributions. Since our purpose is simply to illustrate the multipole structure, we computed only the naive $\mathsf T$-even contribution in the LFCQM. The naive $\mathsf T$-odd contributions have then been obtained by changing the basic multipoles while keeping the same coefficient functions as in the naive $\mathsf T$-even contributions. Note also that the global sign of these naive $\mathsf T$-odd contributions has been chosen arbitrarily. Only a proper calculation including initial and/or final-state interactions can determine this sign.

\subsection{Unpolarized quark in unpolarized target}

\begin{figure}[t]
\centerline{\includegraphics[width=5cm]{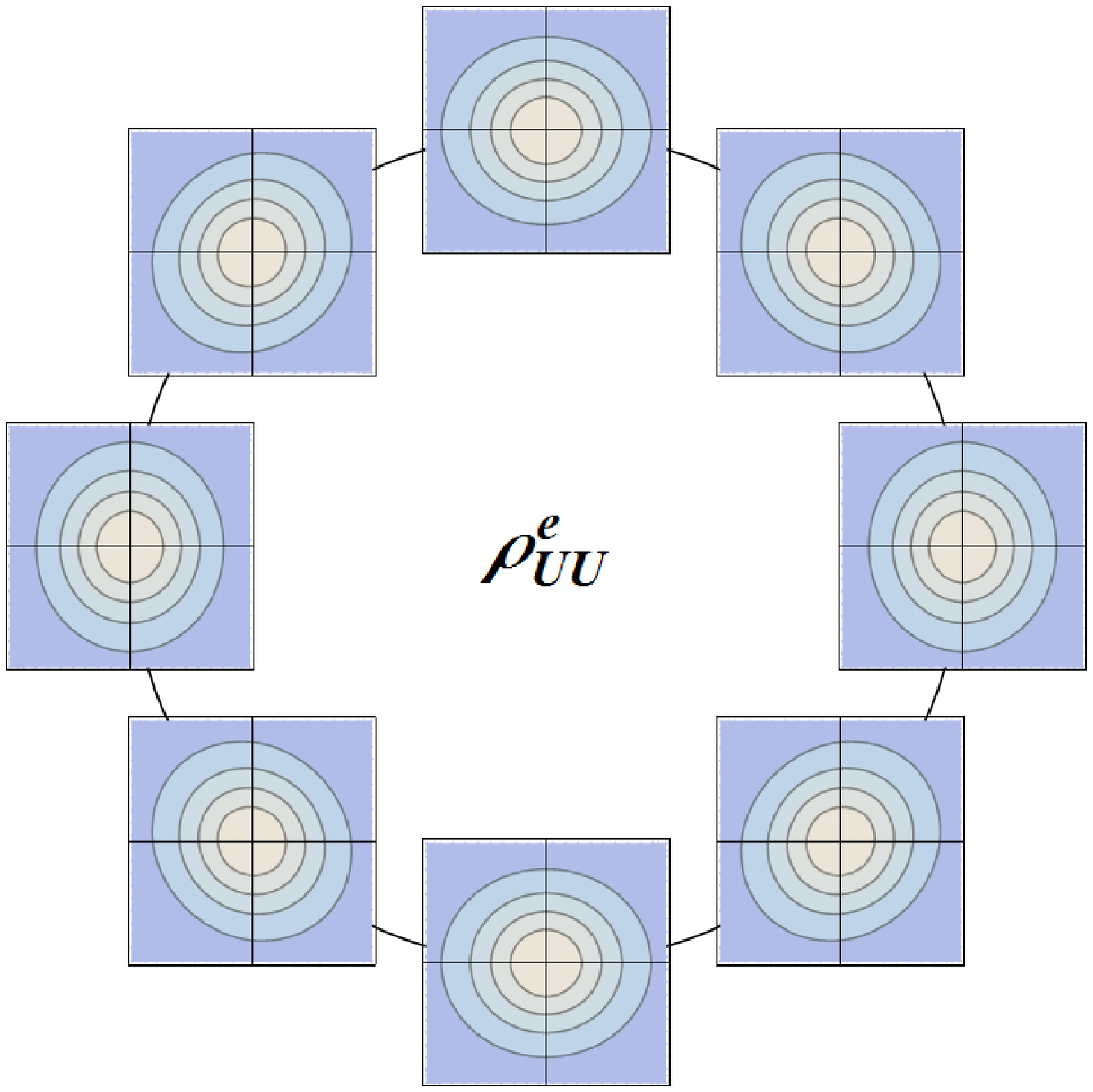}\hspace{1cm}\includegraphics[width=5cm]{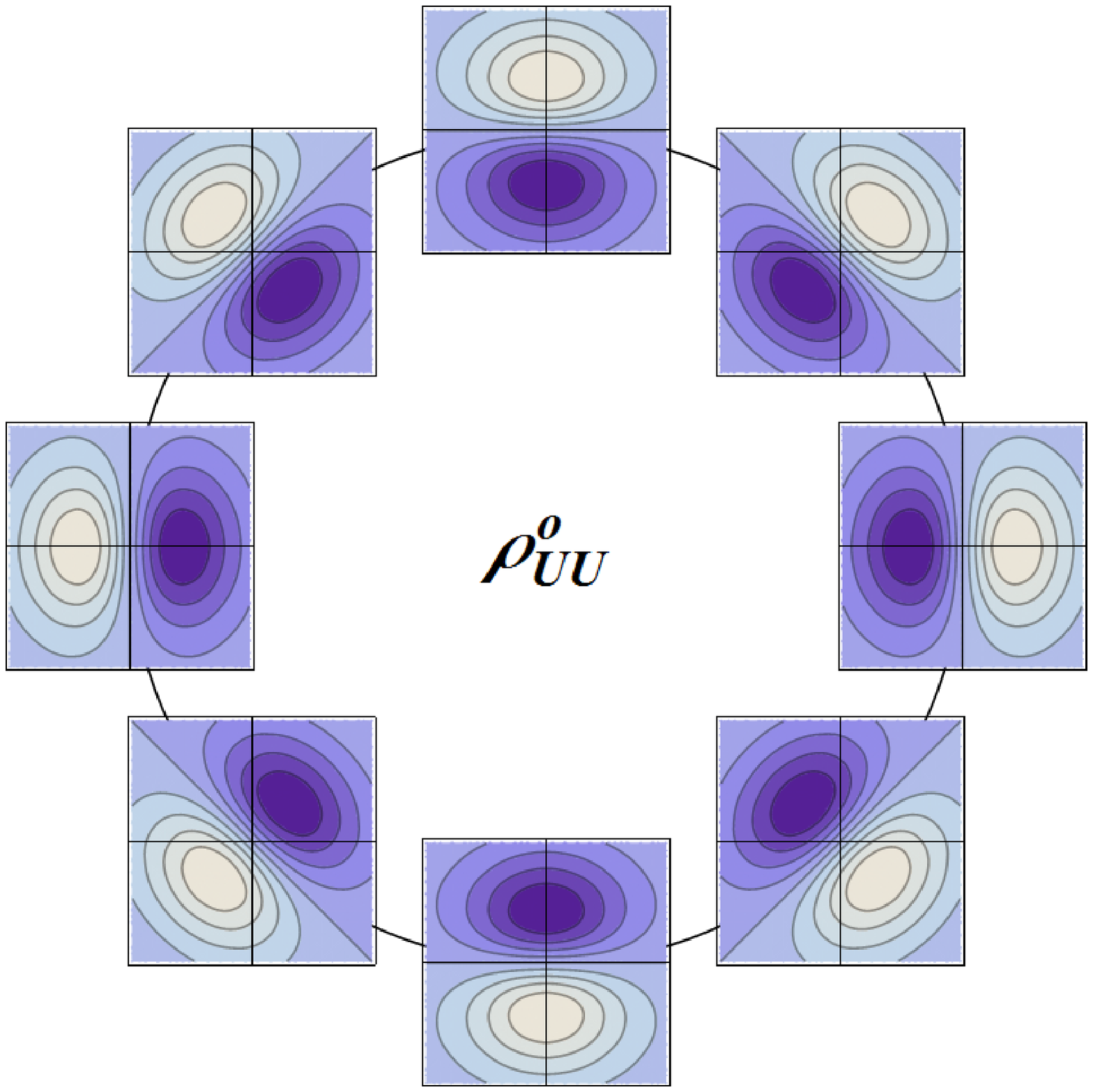}}
\vspace*{8pt}
\caption{Naive $\mathsf T$-even and $\mathsf T$-odd contributions to the transverse Wigner distribution $\rho_{UU}$. See text for more details. \label{fig1}}
\end{figure}

The simplest distribution is $\rho_{UU}$ which describes the distribution of unpolarized quarks in an unpolarized target. In this case, the available transverse vectors are just $\uvec k_T$ and $\uvec b_T$. We then find only two possible basic multipoles
\begin{align}
B^e_{UU}(\uvec k_T,\uvec b_T;\eta)&=1,\\
B^o_{UU}(\uvec k_T,\uvec b_T;\eta)&=\eta\,(\uvec k_T\cdot\uvec b_T).
\end{align}
These two contributions to $\rho_{UU}$ are represented in Fig.~\ref{fig1}. Clearly, the $\rho^o_{UU}$ contribution describes a net inward or outward flow of quarks, which has to be zero for a stable target. A non-vanishing flow therefore originates purely from initial and/or final-state interactions. The coefficient function $C^o_{UU}$ then represents in some sense the strength of the spin-independent part of the attractive or repulsive force associated with initial and final-state interactions.

\subsection{Unpolarized quark in longitudinally polarized target}

\begin{figure}[t]
\centerline{\includegraphics[width=5cm]{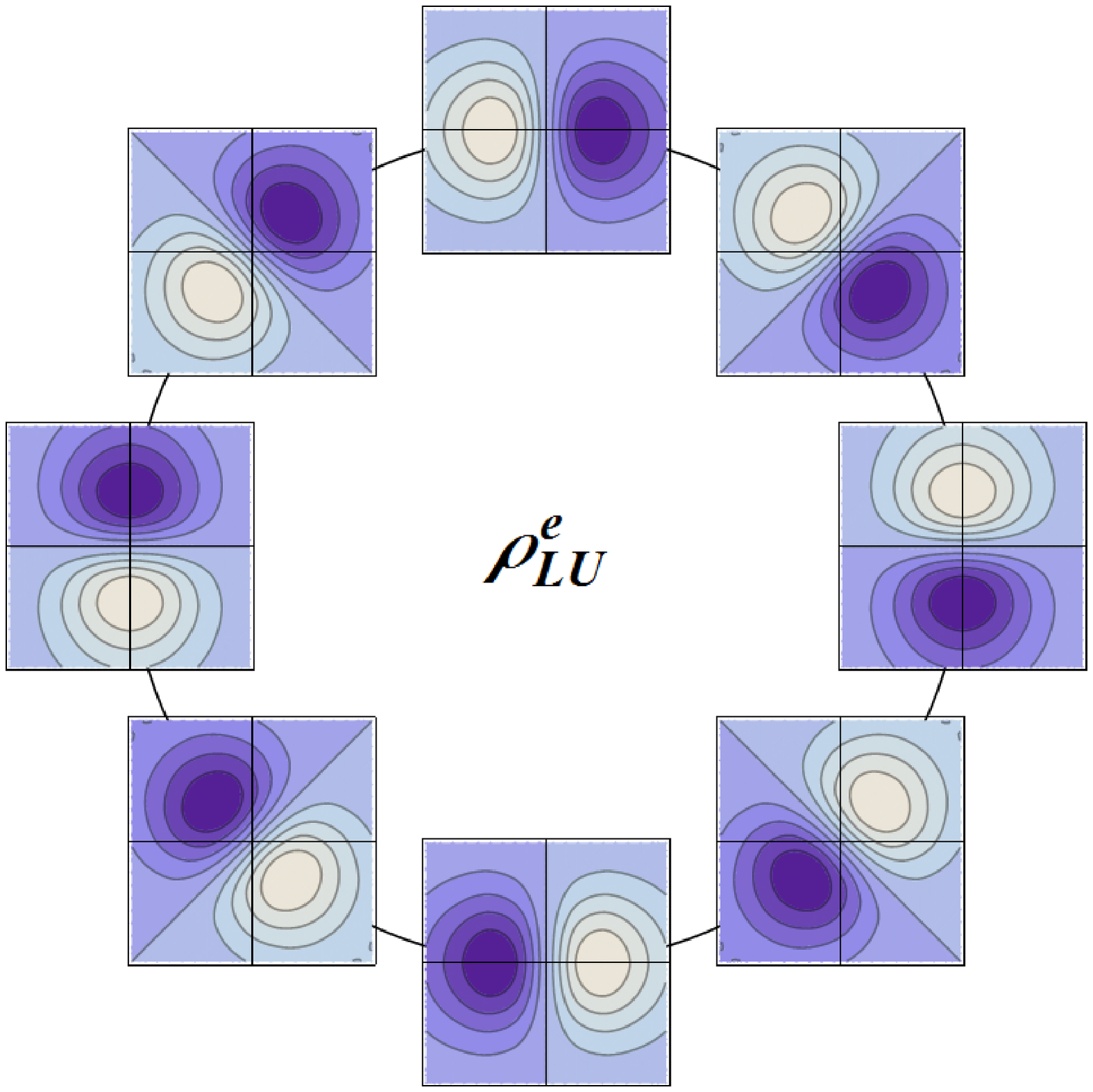}\hspace{1cm}\includegraphics[width=5cm]{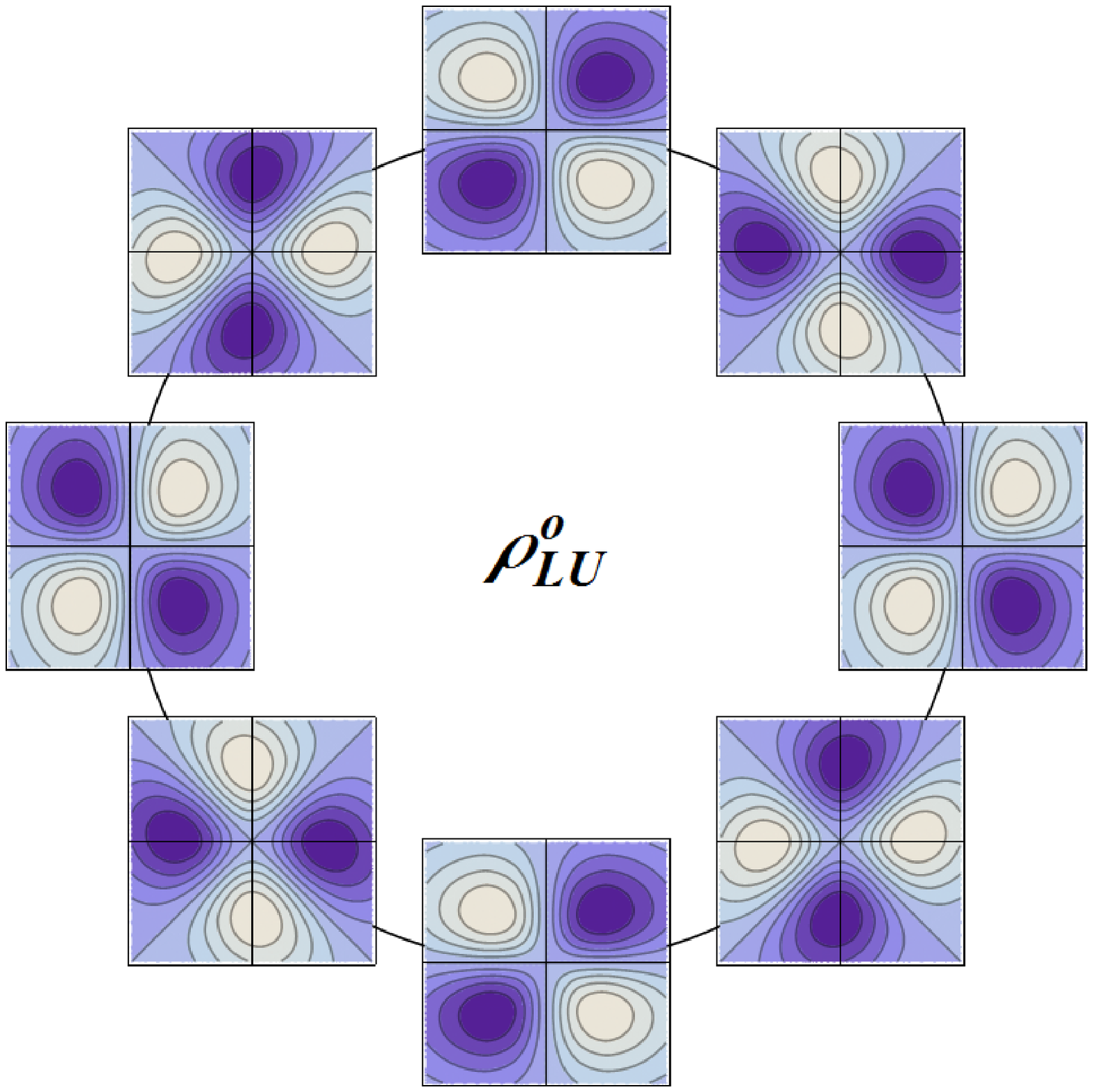}}
\vspace*{8pt}
\caption{Naive $\mathsf T$-even and $\mathsf T$-odd contributions to the transverse Wigner distribution $\rho_{LU}$. See text for more details. \label{fig2}}
\end{figure}

The distribution $\rho_{LU}$ describes the distortion in the distribution of unpolarized quarks due to the longitudinal polarization of the target. We also find only two possible basic multipoles
\begin{align}
B^e_{LU}(\uvec k_T,\uvec b_T;\eta)&=S_L\,(\uvec b_T\times\uvec k_T)_L,\\
B^o_{LU}(\uvec k_T,\uvec b_T;\eta)&=\eta\,S_L\,(\uvec b_T\times\uvec k_T)_L\,(\uvec k_T\cdot\uvec b_T),
\end{align}
where $(\uvec b_T\times\uvec k_T)_L\equiv(\vec b\times\vec k)\cdot \vec P/|\vec P|$. These two contributions to $\rho_{LU}$ are represented in Fig.~\ref{fig2}. Manifestly, the $\rho^e_{LU}$ contribution describes a net circulation of quarks and is therefore directly related to the quark OAM\cite{Lorce:2011kd,Lorce:2011ni}
\begin{equation}
L^q_z=\int\ud x\,\ud^2k_T\,\ud^2b_T\,(\uvec b_T\times\uvec k_T)_L\,\rho_{LU}(x,\uvec k_T,\uvec b_T;\eta).
\end{equation}
Since we worked here with a staple-like gauge link, the resulting $\rho_{LU}$ is related to the canonical version of quark OAM\cite{Lorce:2011kd}. It is also interesting to note that Fig.~\ref{fig2} clearly shows that $\rho^o_{LU}$ cannot contribute to the quark OAM, and hence that the quark OAM is $\eta$-independent\cite{Lorce:2012ce,Hatta:2011ku,Ji:2012ba}.

\subsection{Longitudinally polarized quark in unpolarized target}

\begin{figure}[t]
\centerline{\includegraphics[width=5cm]{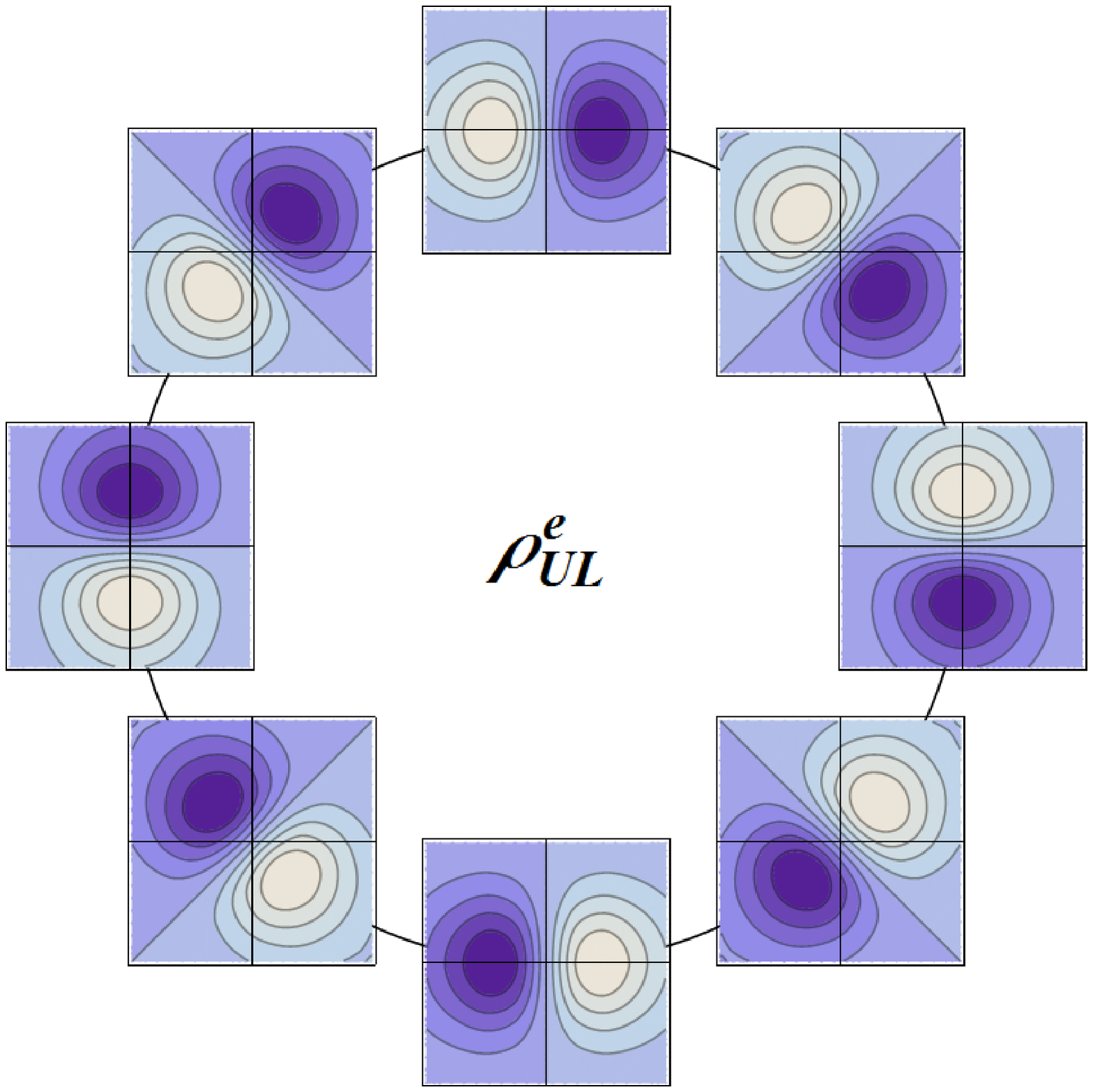}\hspace{1cm}\includegraphics[width=5cm]{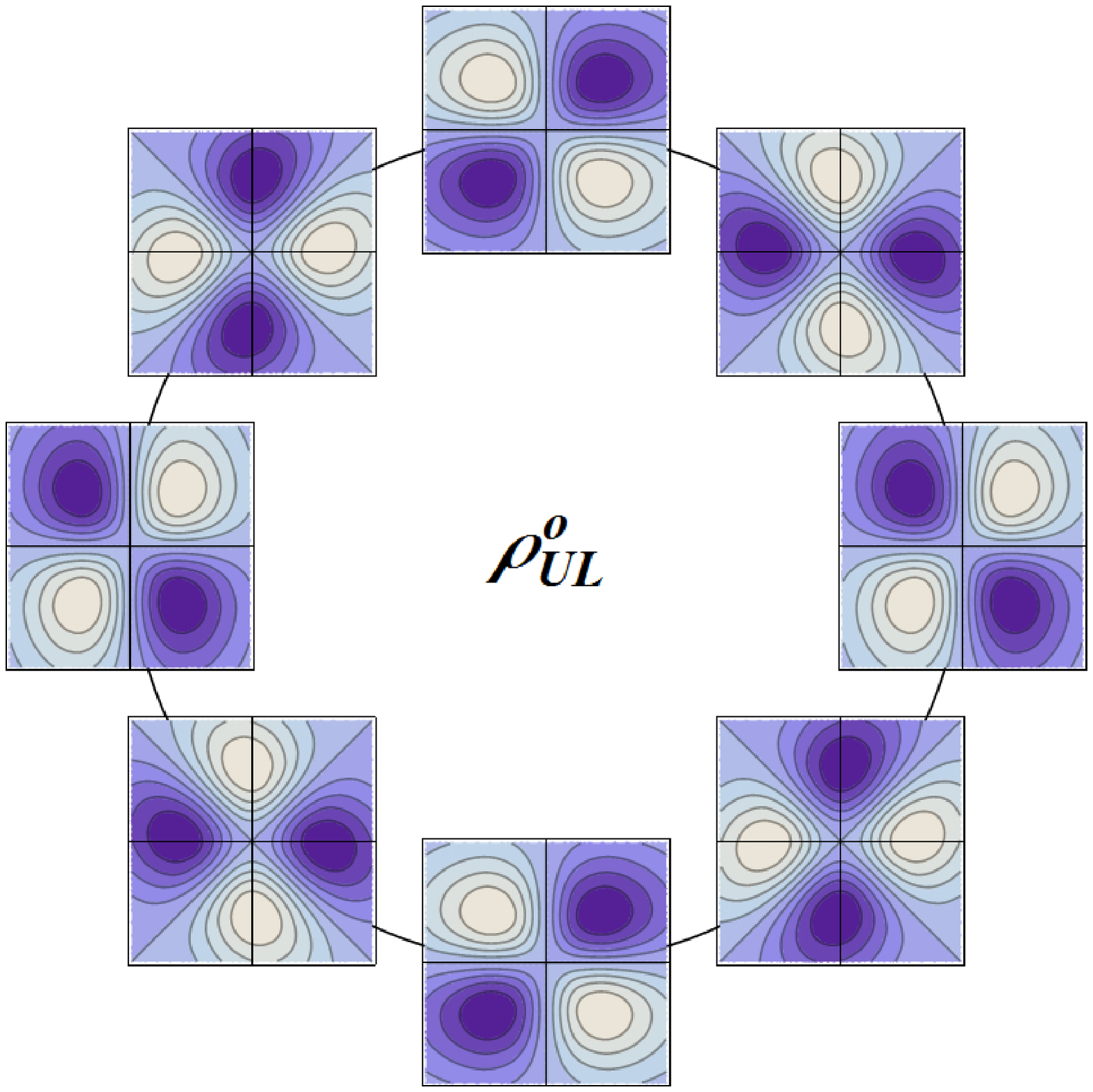}}
\vspace*{8pt}
\caption{Naive $\mathsf T$-even and $\mathsf T$-odd contributions to the transverse Wigner distribution $\rho_{UL}$. See text for more details. \label{fig3}}
\end{figure}

The distribution $\rho_{UL}$ describes the distortion in the distribution of quarks inside an unpolarized target due to the quark longitudinal polarization. This case is very similar to $\rho_{LU}$ since it suffices to replace the target polarization by the quark polarization. We then have
\begin{align}
B^e_{UL}(\uvec k_T,\uvec b_T;\eta)&=S^q_L\,(\uvec b_T\times\uvec k_T)_L,\\
B^o_{UL}(\uvec k_T,\uvec b_T;\eta)&=\eta\,S^q_L\,(\uvec b_T\times\uvec k_T)_L\,(\uvec k_T\cdot\uvec b_T).
\end{align}
These two contributions to $\rho_{UL}$ are represented in Fig.~\ref{fig3}. The $\rho^e_{LU}$ contribution is directly related to the quark spin-orbit correlation\cite{Lorce:2014mxa,Lorce:2011kd}
\begin{equation}
C^q_z=\int\ud x\,\ud^2k_T\,\ud^2b_T\,(\uvec b_T\times\uvec k_T)_L\,\rho_{UL}(x,\uvec k_T,\uvec b_T;\eta).
\end{equation}
Moreover, Fig.~\ref{fig3} clearly shows that the quark spin-orbit correlation is $\eta$-independent. Interestingly, it turns out that in the LFCQM the canonical and kinetic versions of the quark spin-orbit correlation have opposite signs. 

\section{Conclusions}\label{sec5}

We defined the quark spin-orbit correlation and showed how it is connected to parton distributions. This is a new independent piece of information about the nucleon spin structure. Phenomenological estimates indicate that the quark spin and kinetic orbital angular momentum are, in average, opposite. All the various angular correlations can conveniently be seen from a phase-space perspective. We discussed the multipole decomposition pattern and illustrated it with some selected examples obtained from a relativistic quark model.

\section*{Acknowledgements}

I am most grateful to Barbara Pasquini for continuous collaboration. This work was supported by the Belgian Fund F.R.S.-FNRS \emph{via} the contract of Charg\'e de recherches.

\end{document}